\documentclass[twocolumn,secnumarabic,assume, nobibnotes, aps, prd]{revtex4-2}

\usepackage{amsmath}
\usepackage{amsfonts}
\usepackage{amssymb}

\usepackage{psfrag,graphicx}
\usepackage{float}
\usepackage[T1]{fontenc}
\usepackage{xcolor}
\usepackage{lmodern}
\usepackage{listings}
\usepackage{subfigure}
\usepackage[section]{placeins}
\usepackage{psfrag}
\usepackage{rotating}
\usepackage{multirow}
\usepackage{colortbl}
\allowdisplaybreaks
\usepackage{parskip}
\usepackage[T1]{fontenc}
\begin{document} 
\title{Unidirectional flow of flat-top solitons}%

\author{M. O. D. Alotaibi$^{1}$, L. Al Sakkaf$^{2}$, and U. Al Khawaja$^{2,3}$}
\address{$^1$Department of Physics, Kuwait University, P.O. Box 5969 Safat, 13060 Kuwait\\$^2$Department of Physics, United Arab Emirates University, P.O. Box 15551, Al-Ain, United Arab Emirates\\$^3$Department of Physics, School of Science, The University of Jordan, Amman, 11942, Jordan}
%

\begin{abstract}
We numerically demonstrate the unidirectional flow of flat-top solitons when interacting with two reflectionless potential wells with slightly different depths. The system is described by a nonlinear Schr\"{o}dinger equation with dual nonlinearity. The results show that for shallow potential wells, the velocity window for unidirectional flow is larger than for deeper potential wells. A wider ﬂat-top solitons also have a narrow velocity window for unidirectional ﬂow than those for thinner flat-top solitons.
\end{abstract}

\maketitle


\section{Introduction}
\label{sec:Introduction}
Solitons are a class of nonlinear waves that demonstrate remarkable stability properties as a result of the balance between non-linear and dispersive effects within the medium~\cite{book1}. They are found in a wide range of physics, including hydrodynamic tsunamis~\cite{1a}, fiber optic communications~\cite{1b}, solid-state physics~\cite{1c}, and the dynamics of biological molecules~\cite{1d}. In addition, solitons in optics can exhibit unidirectional flow when interacting with a specific type of potential~\cite{2a,2b,2c}. The significance of the unidirectional flow appealed to both fundamental scientists and applied scientists interested in developing new technologies related to soliton~\cite{3a,3b,3c,3d,3e,3f,3g,3h,3i,3j}. A novel type of soliton is the flat-top soliton (FTS), characterized by a flat-top profile with a finite width and a sharp edge~\cite{FTS1,FTS2,FTS3,FTS4,FTS5,FTS6,FTS7,FTS8,FTS9,FTS10,FTS11}. It is a solution to the nonlinear Schr\"{o}dinger equations (NLSE) with dual nonlinearity. The flat top shape is due to the fact that the competition between the cubic and quintic nonlinear components in the NLSE imposes an upper limit on the wave field density. Therefore, an increase in the total norm of the field causes the width to expand while the amplitude remains constant. They have been observed in various physical systems, such as Bose-Einstein condensates~\cite{FTS12}, optical fibers~\cite{FTS13}, and microresonators~\cite{FTS14}, and have potential applications in optical communication, all-optical switching, and frequency comb generation.

In this article, we study the interaction of FTS with a particular class of potentials called P\"{o}schl-Teller potential~\cite{PoschlTeller1,PoschlTeller2}. A remarkable property of P\"{o}schl-Teller potentials is that they are "reflectionless," which means that the scattering of FTS characterized by the absence of radiation. By considering a double-well reflectionless potential of this type with slightly varying depths, we demonstrate numerically that a unidirectional flow of FTS can be achieved. To the best of our knowledge, the unidirectional flow of FTS has not been previously investigated. In order to confirm our findings, we performed the numerical calculations using two different methods, the split-step method, and the power series expansion method~\cite{Sakkaf1a}. Both methods show similar results. 

We followed the calculations in Ref.~\cite{Sakkaf1b}, where a single parameter, $\gamma$, controls the FTS width. Depending on this value, we obtain a bright soliton or FTSs with different widths. Next, we select parameters for the potential that result in a relatively large velocity window for the unidirectional flow of a bright soliton. Then, we fix the parameters of the potential and vary $\gamma$ to obtain FTSs and investigate their unidirectional flow. 

The paper is structured as follows. In Sec.~\ref{sec:setup_sec}, we present our model and define the problem. In Sec.~\ref{sec:Numerical_Results}, we perform numerical simulations to demonstrate that the unidirectional flow of FTS can be accomplished when it interacts with a reflectionless asymmetric double-well potential. In addition, we investigate the behavior of the velocity window for unidirectional flow as the FTS width and potential depth is altered. Finally, Sec.~\ref{sec:Conclusions} concludes with a summary of our findings. 


\section{Setup and theoretical model}
\label{sec:setup_sec}

The following dimensionless NLSE with cubic and quintic nonlinearity terms describes the dynamics of FTS,

\begin{align}
	\label{eq:EOMFTS}
&i \frac{\partial}{\partial t} \psi\left(x,t\right)  +g_{1} \frac{\partial^2}{\partial x^2} \psi\left(x,t\right)  + g_{2} |\psi\left(x,t\right)|^{2} \psi\left(x,t\right) \nonumber  \\ 
& + g_{3} |\psi\left(x,t\right)|^{4} \psi\left(x,t\right) + V(x) \psi\left(x,t\right)=0, 
\end{align}

where $\psi\left(x,t\right)$ is a complex field, $g_{1}$ represents the strength of dispersion, $g_{2}$, and $g_{3}$ characterize the strengths of the two nonlinearity terms, and $V(x)$ is an external potential. The solution to Eq.~\eqref{eq:EOMFTS} has the following form,

\begin{align}
	\label{eq:Sol_FTS}
	\nonumber 
&\psi\left(x,t\right) = \sqrt{\frac{2u_{0}}{g_{2}\sqrt{1+\gamma}}} \\ & \times
\frac{1}{\sqrt{\frac{1-\sqrt{1+\gamma}}{2\sqrt{1+\gamma}}+\mathrm{cosh}^2\left[\sqrt{\frac{u_{0}}{g_{1}}}(x-x_{0}-v_{0}t)\right]}}  e^{i \phi\left(x,t\right)},
\end{align}

where $u_{0}$, $x_{0}$ and $v_{0}$ are arbitrary parameters that represent the amplitude, peak position, and soliton velocity, respectively. The phase is $\phi\left(x,t\right) = u_{0} t + v_{0}\left[2\left(x-x_{0}\right)-v_{0}t\right]/\left(4 g_{1}\right)$. The parameter $\gamma = g_{3}/g_{30}$, where $g_{30} = 3 g^2_{2}/16 u_{0}$, determines the solution profile. Depending on the value of $\gamma$, we may obtain a bright soliton, kink soliton, thin-top soliton, or FTS~\cite{Sakkaf1b}. 

In order to obtain FTS, $\gamma$ should be in the range of $-1 < \gamma <0$. For $\gamma=0$, the solution reduces to the bright soliton. As $\gamma$ approaches -1, the width of the FTS increases such that at $\gamma=-1$, a kink soliton is obtained, which may be regarded as a FTS with infinite width. It is thus convenient to adjust the width of the FTS using $\gamma$ according to $\gamma=\gamma_{i} = 10^{-i}-1$ such that for $i=0$, we obtain a bright soliton, $\gamma_{0}=0$. In the numerical simulations, the machine precision sets a maximum limit on the width of the FTS. For $i=16$, we acquire the widest FTS that can be simulated~\cite{Sakkaf1b}. The potential, $V(x)$, in Eq.~\eqref{eq:EOMFTS} is an asymmetric double-well potential,
\begin{align}
	\label{eq:potential}
	V\left(x\right) = V_{1} \mathrm{sech}^2\left[\alpha_{1}\left(x-\beta\right)\right] +  V_{2} \mathrm{sech}^2\left[\alpha_{2}\left(x+\beta\right)\right] ,
\end{align}
where $V_{1,2}$, $\alpha_{1,2}$, and $\beta$ are the potential depths, inverse widths, and locations of the double-well, respectively.

\begin{figure}[!h]    
	\centering
\includegraphics[width=\columnwidth]{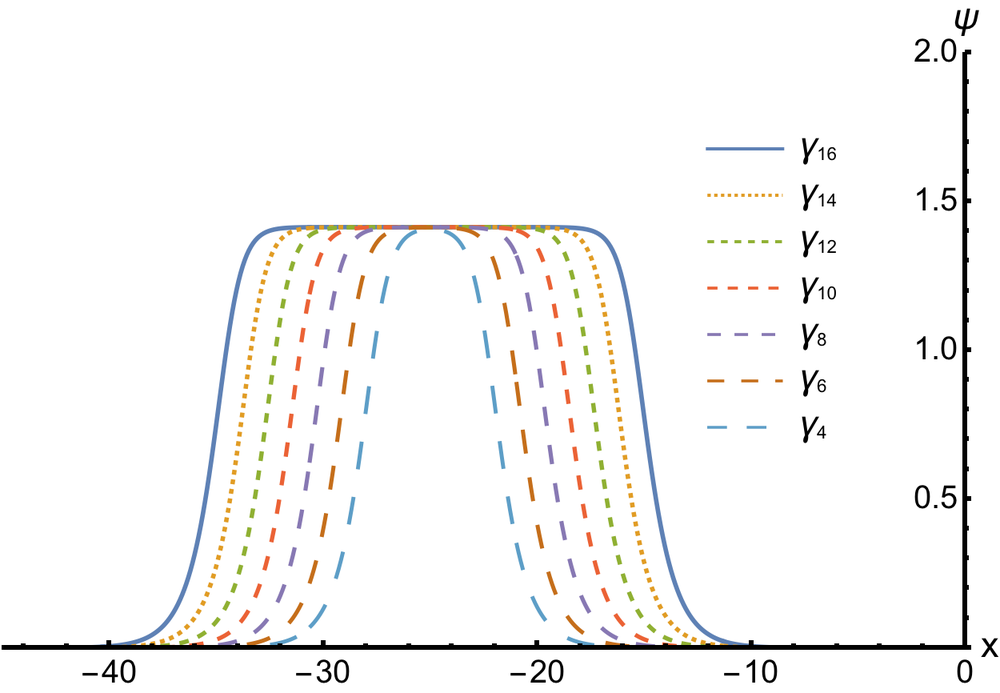}
	\caption{Flat-top soliton profiles for different values of $\gamma_{i}$ in Eq.~\eqref{eq:Sol_FTS}. Here, $\gamma_{i} = 10^{-i}-1$. Other parameters are $u_{0}=0.5$, $ g_{1}=0.5$, $ g_{2}=1$, $ x_{0}=-25$, and $v_{0}=0$.} 
	\label{fig:FRHPRA:FTS_densities}
\end{figure} 

\begin{figure}[!h]    
	\centering
\includegraphics[width=\columnwidth]{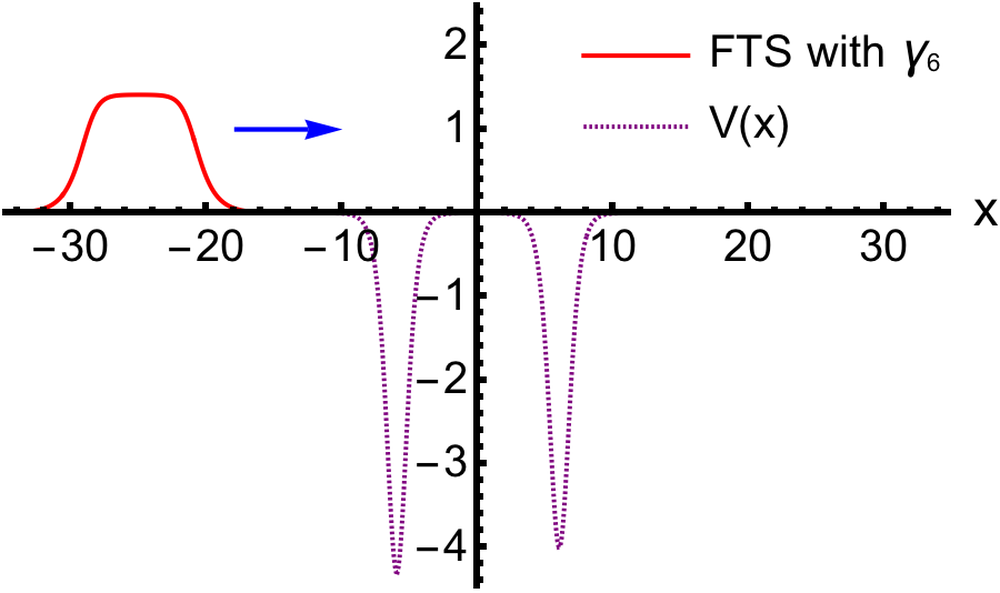}
	\caption{Flat-top soliton profile with a width corresponding to $\gamma_{6}$ and a double-well potential. The asymmetric double-well potential parameters in Eq.~\eqref{eq:potential} are $V_{1}=-4$, $V_{2}=-4.3$, $\alpha_{1}=\alpha_{2}=1$, and $\beta=6$. We use these configurations in the paper to study the unidirectional propagation of the FTS. The arrow shows the situation for a FTS approaching from the left side.} 
	\label{fig:FRHPRA:potentialProfile}
\end{figure}

In Fig.~\ref{fig:FRHPRA:FTS_densities}, We plot the FTS profile for various $\gamma$ values. In the rest of the paper, we use a FTS with a width corresponding to $\gamma_{6}$ unless stated otherwise. The potential parameters, as well as the FTS used in the numerical calculations, are depicted in Fig.~\ref{fig:FRHPRA:potentialProfile}.

Following the conventional rule of obtaining the results from two different approaches, we use two numerical methods, namely the split-step method and the power series expansion method~\cite{Sakkaf1a}, to confirm the unidirectional flow of the FTS. Both methods give similar results. 


 
\section{Numerical Results}
\label{sec:Numerical_Results}
To study the unidirectional flow of the FTS, the potential is fixed at the center, the FTS is launched from both sides, and the scattered region is thus observed. We set the FTS in motion with the initial center-of-mass velocity, $v_{0}$, toward the potential area, then calculate the reflection (R), transmission (T), or trapping (L) coefficients which are known as transport coefficients. When we send the FTS from left to right, the transport coefficients are defined as,
\begin{align}
	\label{eq:RLT}
R =& \frac{1}{N}\int_{-\infty}^{-\delta} |\psi\left(x,t\right)|^2 dx, \\ \nonumber
L  =& \frac{1}{N}\int_{-\delta}^{\delta}  |\psi\left(x,t\right)|^2 dx, \\ \nonumber
T =& \frac{1}{N}\int_{\delta}^{\infty}   |\psi\left(x,t\right)|^2 dx,
\end{align}
where $\delta$ is the position of measurement of reflectance or transmission, set at a value slightly greater than the position of the potential boundary and $N = \int_{-\infty}^{\infty} |\psi|^2 dx$ is the normalization of the FTS. The $R$ and $T$ exchange roles if we send the FTS from right to left. The three coefficients must satisfy the conservation law $R + T + L = 1$. 

\begin{figure}[!h]
	\centering
\includegraphics[width=\columnwidth]{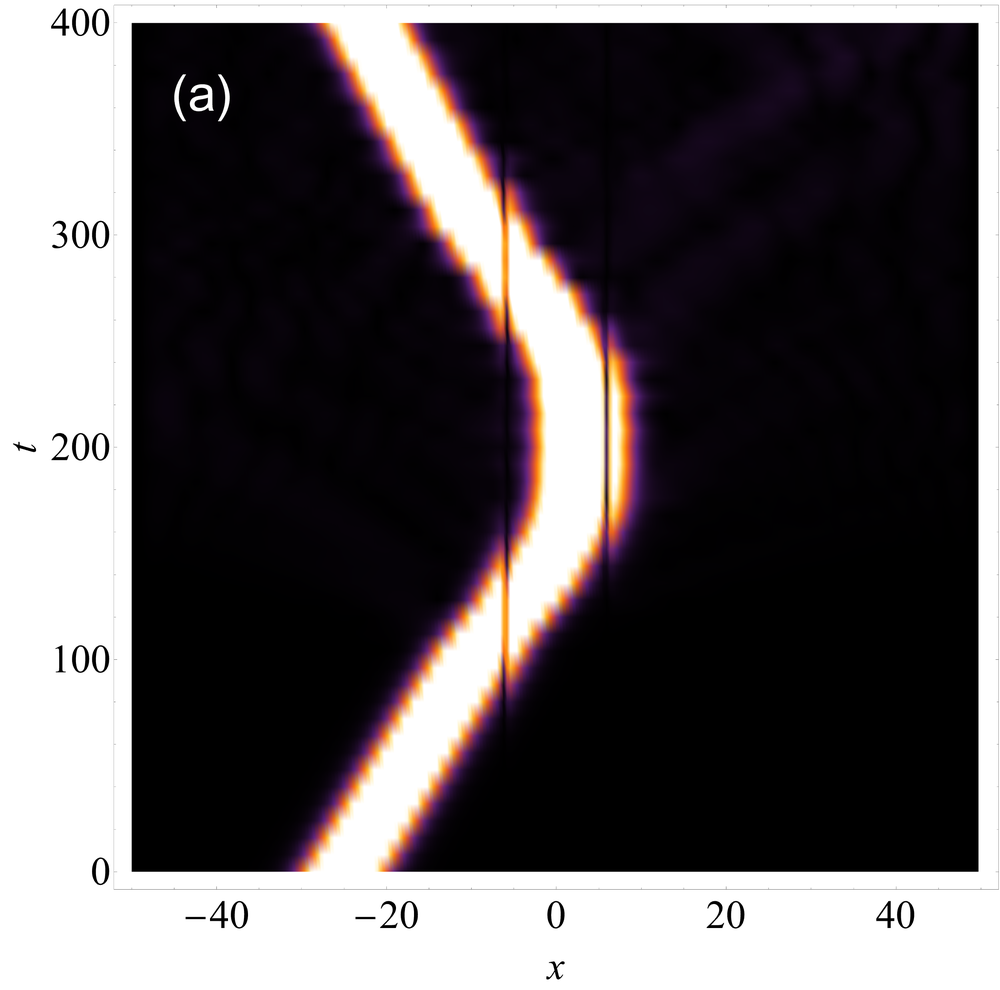}
\includegraphics[width=\columnwidth]{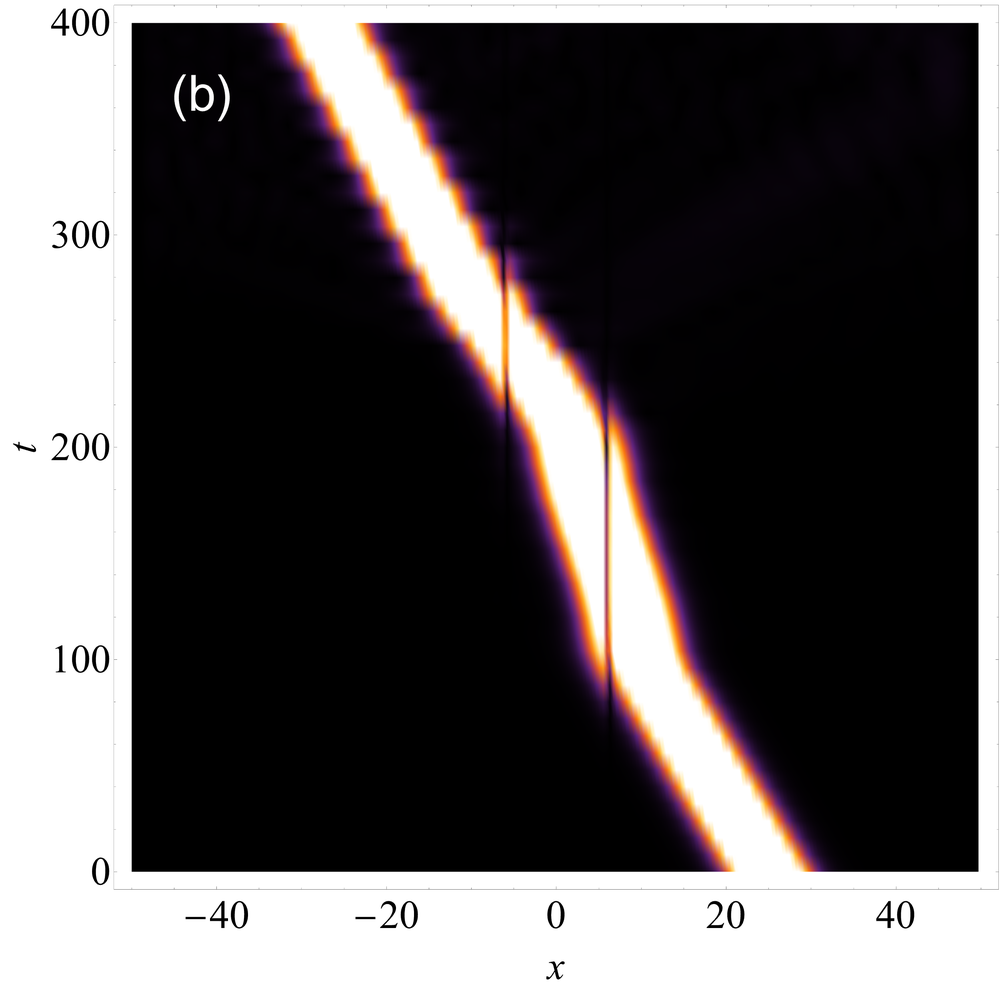}
	\caption{Density plot of a unidirectional flow of a FTS interacting with an asymmetric double-well potential. In (a), the FTS set in motion from $x_{0}=-25$ with velocity $v_{0}=0.15$, and in (b), the FTS set in motion from $x_{0}=25$ with velocity $v_{0}=-0.15$. We use a FTS with a width corresponding to $\gamma_{6}$ and potential parameters $V_{1}=-4$, $V_{2}=-4.3$, $\alpha_{1}=\alpha_{2}=1$, and $\beta=6$.}
	\label{fig:FRHPRA:uni_dir_1_FTS}
\end{figure}

In Fig.~\ref{fig:FRHPRA:uni_dir_1_FTS}, we plot an FTS with a width corresponding to $\gamma_{6}$ in Fig.~\ref{fig:FRHPRA:FTS_densities}. We first send the FTS from the left at $x=-25$ with a velocity of $v_{0}=0.15$. As a result, the FTS interacts with the asymmetric potential described in Fig.~\ref{fig:FRHPRA:potentialProfile}, then reflects. But when we send the FTS from the right with $x=25$ and $v_{0}=-0.15$, we find that the FTS transmitted over the potential. The next step is to determine whether this phenomenon occurs throughout a wide range of velocities. We do so by calculating the transport coefficients defined in Eq.~\eqref{eq:RLT}. 

\begin{figure}[!h]
	\centering
\includegraphics[width=\columnwidth]{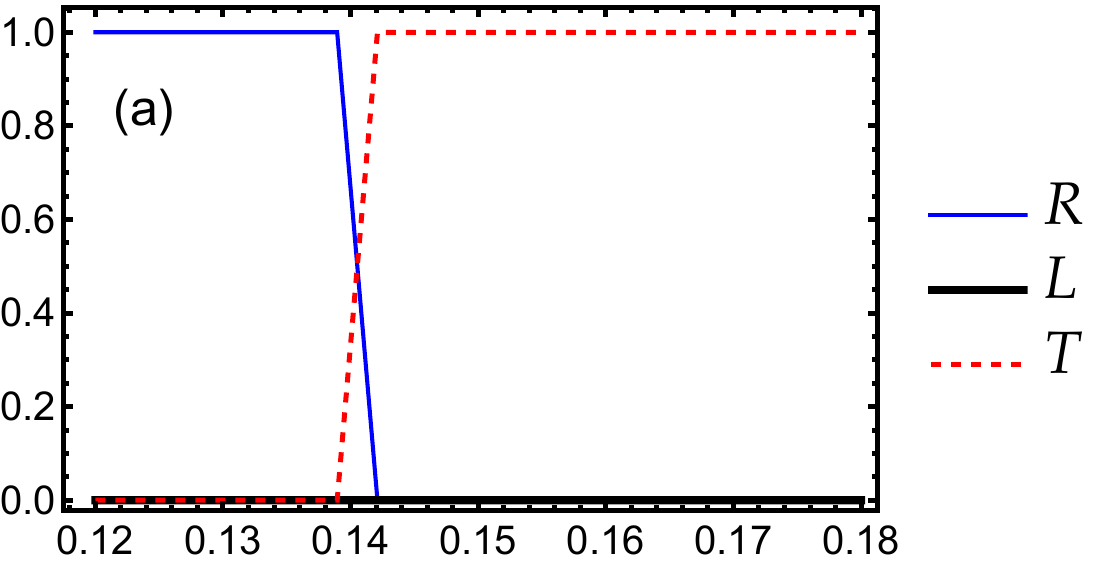}
\includegraphics[width=\columnwidth]{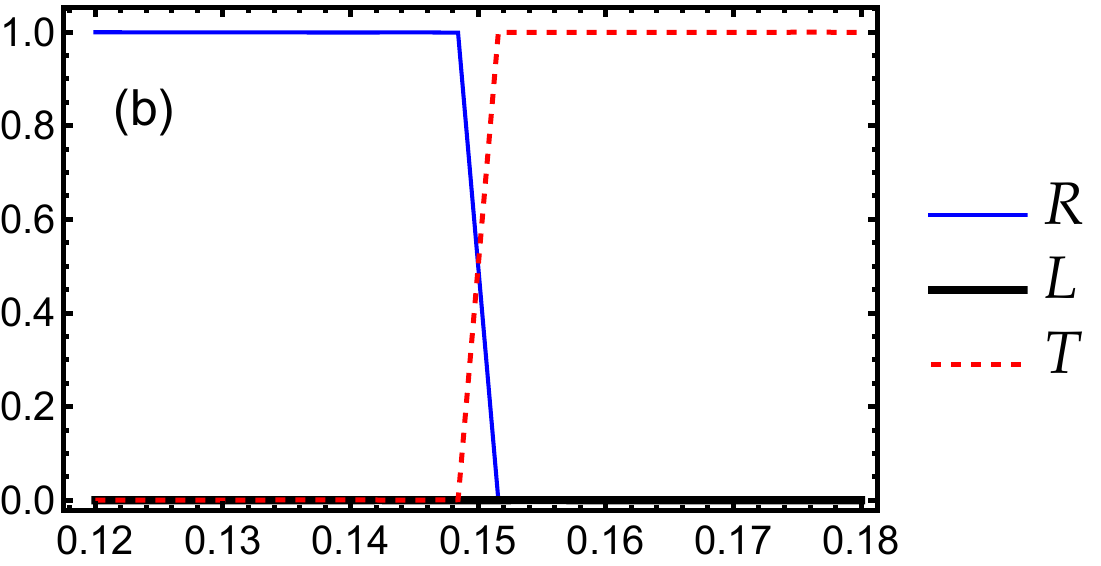}
	\caption{Transport coefficients for the FTS with a width corresponding to $\gamma_{6}$ as shown in Fig.~\ref{fig:FRHPRA:uni_dir_1_FTS}. (a) for a soliton starting from the right side and moving towards the potential well and (b) for a soliton starting from the left side of the potential well. The velocity window range for the unidirectional flow is between $v=0.14$ and $v=0.15$.}
	\label{fig:FRHPRA:TRL_gamma_6}
\end{figure}

In Fig.~\ref{fig:FRHPRA:TRL_gamma_6}, we plot the transport coefficients for the FTS in Fig.~\ref{fig:FRHPRA:uni_dir_1_FTS}. We see that there is a velocity window, $0.14<v_{0}<0.15$, where the unidirectional flow exists. Depending on the FTS width, we may have different velocity windows for the unidirectional flow. Therefore, it is instructive to calculate the transport coefficients for a wide range of FTS widths. 

\begin{figure}[!h]
	\centering
\includegraphics[width=\columnwidth]{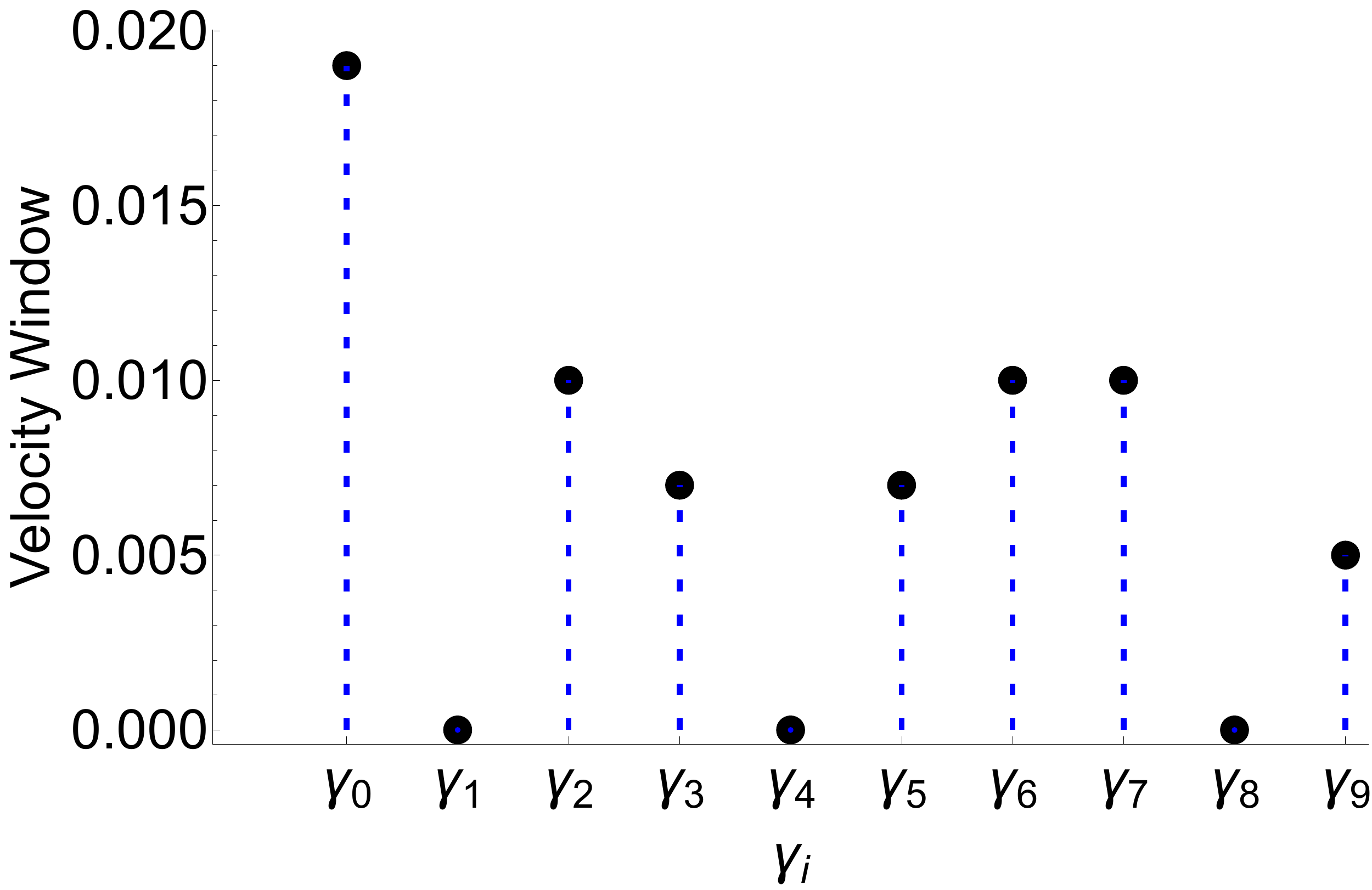}
	\caption{Velocity window for a unidirectional flow of FTS with different $\gamma_{i}$ (width). Here, $\gamma_{i} = 10^{-i}-1$ such that $\gamma_{0} = 0$ corresponds to a bright soliton. We use potential parameters similar to the ones in Fig.~\ref{fig:FRHPRA:uni_dir_1_FTS}.}
	\label{fig:FRHPRA:SolitonWidth_vs_VW}
\end{figure}

Figure~\ref{fig:FRHPRA:SolitonWidth_vs_VW} shows the velocity window for a unidirectional flow of the FTS with varying widths but with potential parameters similar to the ones in Fig.~\ref{fig:FRHPRA:potentialProfile}. We find the velocity window for unidirectional flow equal to $0.019$ starting from bright soliton, $\gamma_{0}$. The velocity window shrinks as we raise the width until we reach a FTS with a width of $\gamma_{9}$. We found that FTS with a width of $\gamma_{10}$ and higher breaks into small parts. As a result, the velocity window is not calculated for these values. 

\begin{figure}[!h]
	\centering
\includegraphics[width=\columnwidth]{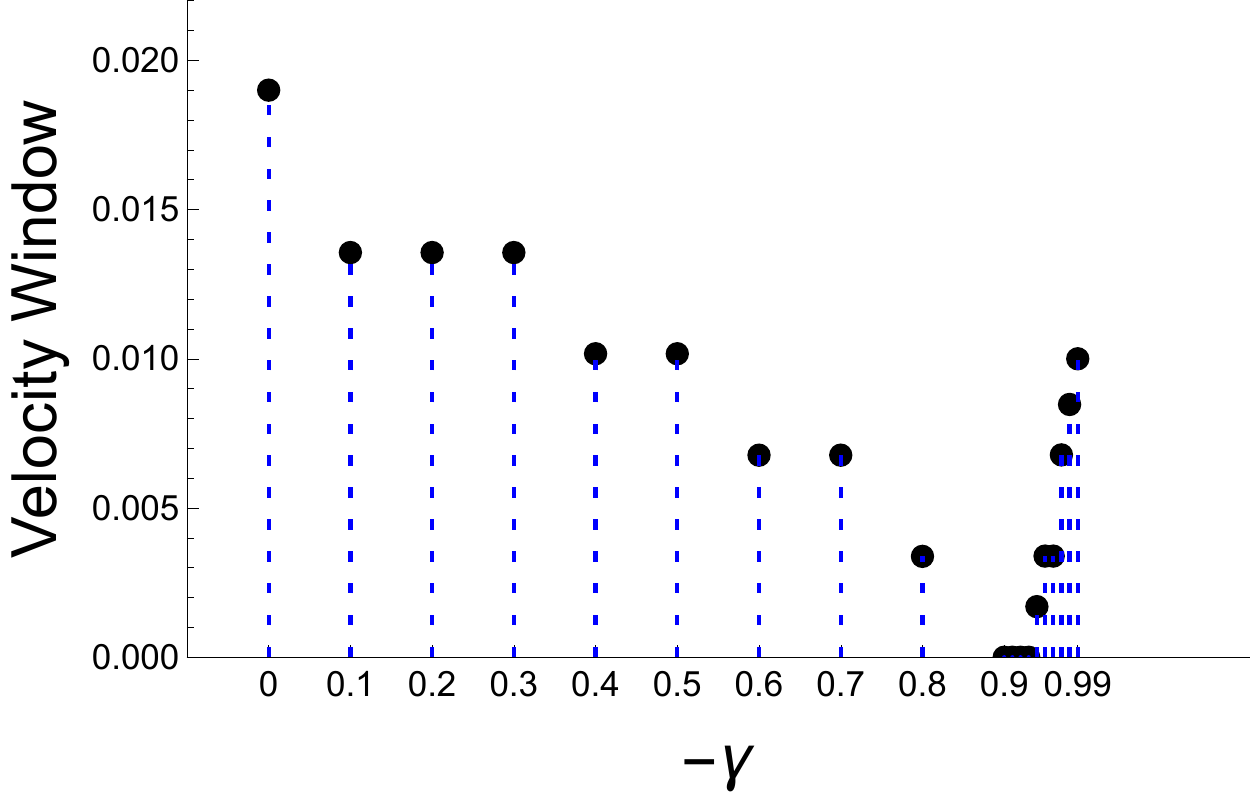}
	\caption{Velocity window for a unidirectional flow of FTS for widths between $\gamma_{0}=0$ and $\gamma_{2}=0.99$ in Fig.~\ref{fig:FRHPRA:SolitonWidth_vs_VW}. The velocity window for the unidirectional flow at $\gamma_{1}=0.9$ disappears and reappears when moving in either direction away from $\gamma_{1}$.}
	\label{fig:FRHPRA:SolitonWidth_gamma0_gamma2}
\end{figure}

A noteworthy observation is that the unidirectional flow ceases to exist for specific FTSs, namely $\gamma_{1}$, $\gamma_{4}$, and $\gamma_{8}$ as shown in Fig.~\ref{fig:FRHPRA:SolitonWidth_vs_VW}. According to the results, there is a decrease in the velocity window as the width increases from $\gamma_{0}$ to $\gamma_{1}$. Eventually, the velocity window disappears completely at $\gamma_{1}$, and then there is an increase in proximity to $\gamma_{2}$, as illustrated in Fig.~\ref{fig:FRHPRA:SolitonWidth_gamma0_gamma2}. Similar observations occur around $\gamma_{4}$ and $\gamma_{8}$. The existence of a velocity window for unidirectional flow is a well-established phenomenon, typically confined to a specific parameter space. Deviation from this region results in the cessation of unidirectional flow. The noteworthy observation is that the unidirectional flow characteristic resurfaces upon transitioning to a distinct parameter domain. A comprehensive examination could potentially elucidate the underlying physics of this phenomenon, which remains a topic for future investigation.  

\begin{figure}[!h]
	\centering
\includegraphics[width=\columnwidth]{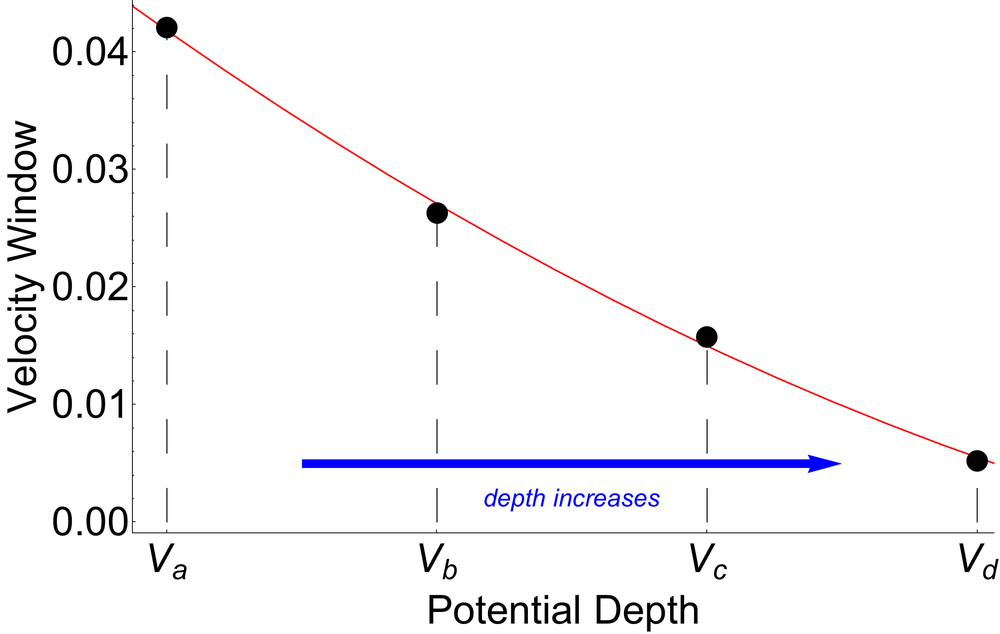}
	\caption{Velocity window for a unidirectional flow of FTS with width corresponds to $\gamma_{3}$. The potential parameters are similar to those given in Fig.~\ref{fig:FRHPRA:uni_dir_1_FTS}, but with $V_{1,2} = -1,-1.3$ in $V_{a}$, $V_{1,2} = -2,-2.3$ in $V_{b}$, $V_{1,2} = -3,-3.3$ in $V_{c}$ and $V_{1,2} = -4,-4.3$ in $V_{d}$ }
	\label{fig:FRHPRA:VWvsPotential}
\end{figure}

In Figure~\ref{fig:FRHPRA:VWvsPotential}, we look at how increasing the potential depth affects the velocity window. We fix the FTS width here by selecting a FTS with a width corresponding to $\gamma_{3}$ and changing the asymmetric double-well potential depth. It is known that only a double-well potential with a slightly different depth can generate the unidirectional flow of a soliton~\cite{2a}. Therefore, we calculate the velocity window for $V_{1,2}=-1, -1.3$ in $V_a$,  $V_{1,2}=-2, -2.3$ in $V_b$,  $V_{1,2}=-3, -3.3$ in $V_c$ and, lastly,  $V_{1,2}=-4, -4.3$ in $V_d$  which is the case in Fig.~\ref{fig:FRHPRA:potentialProfile}. We found that the velocity window for shallow potential wells is larger than for deeper potential wells. The red line in Fig.~\ref{fig:FRHPRA:VWvsPotential} guides the eye and indicates the lowering of the velocity window as we increase the depth.


\section{Conclusions}
\label{sec:Conclusions}

Our numerical analysis of the dynamics of the one-dimensional NLSE with competing cubic-quintic nonlinearity terms and two asymmetric reflectionless type potential wells demonstrates that the FTS can propagate in one direction for a specific velocity range and with specific parameters. To validate our findings, we employed two computational techniques, namely the split-step method and the power series expansion method. Our analysis indicates that both methodologies yield comparable outcomes. In addition, through the establishment of fixed potential parameters and utilization of bright soliton as a reference point, an examination was conducted on the velocity range pertaining to the unidirectional flow of FTSs possessing varying widths. We show that the FTSs have smaller ones for a relatively large unidirectional flow velocity window for the bright soliton. Moreover, by increasing the FTS widths, the velocity window decreases. Furthermore, for specific FTS widths, the unidirectional flow ceases to exist, resulting in symmetrical dynamics regardless of whether the FTS is introduced from the right or left of the potential. Also, upon fixing the width of the FTS and manipulating the depths of the potential, it was observed that the range of unidirectional flow velocities is greater for shallow potential wells in comparison to deeper ones. 

Subsequent research endeavors could expand upon our investigation by conducting an analytical analysis of the unidirectional propagation of FTSs, with the aim of further elucidating the underlying physics responsible for the cessation of unidirectional flow in specific FTS widths.

\section*{Acknowledgments} 
M. O. D. Alotaibi is grateful to the Physics Department at the United Arab Emirates University for their hospitality during his visit.

\end{document}